\documentclass[natbib]{emulateapj}
\usepackage{footnote}
\usepackage{amsmath}
\usepackage{hyperref}
\hypersetup
{
bookmarksopen=true,
pdftitle="The inner edge of the HZ",
pdfauthor="", 
pdfmenubar=true, 
pdfhighlight=/O, 
pdfpagemode=UseNone, 
pdfpagelayout=SinglePage, 
pdffitwindow=true, 
linkcolor=blue,
citecolor=linkcol, 
urlcolor=linkcol 
}

\shorttitle{A population-based Habitable Zone perspective}
\shortauthors{A. Zsom}

\begin{document}

\title{A population-based Habitable Zone perspective}

\author{Andras Zsom}
\affil{Department of Earth, Atmospheric and Planetary Sciences, Massachusetts Institute of Technology, Cambridge, MA 02139, USA; \texttt{zsom@mit.edu}}

\begin{abstract}
What can we tell about exoplanet habitability if currently only the stellar properties, planet radius, and the incoming stellar flux are known? A planet is in the Habitable Zone (HZ) if it harbors liquid water on its surface. The HZ is traditionally conceived as a sharp region around stars because it is calculated for one planet with specific properties. Such an approach is limiting because the planetsÕ atmospheric and geophysical properties, which influence the presence of liquid water on the surface, are currently unknown but expected to be diverse. 

A statistical HZ description is outlined which does not favor one planet type. Instead the stellar and planet properties are treated as random variables and a continuous range of planet scenarios are considered. Various probability density functions are assigned to each random variable, and a combination of Monte Carlo sampling and climate modeling is used to generate synthetic exoplanet populations with known surface climates. Then, the properties of the liquid water bearing subpopulation is analyzed. 

Given our current observational knowledge, the HZ takes the form of a weakly-constrained but smooth probability function. The HZ has an inner edge but a clear outer edge is not seen. Currently only optimistic upper limits can be derived for the potentially observable HZ occurrence rate. Finally, we illustrate through an example how future data on exoplanet atmospheres will help to narrow down the probability that an exoplanet harbors liquid water and we identify the greatest observational challenge in the way of finding a habitable exoplanet.

\end{abstract}


\section{Introduction}
One way to answer the question whether there is life elsewhere in the Universe is to discover rocky exoplanets and to characterize their atmospheres in hope of finding remotely detectable signs of life (biosignatures). This is a challenging task because observations need to be precise enough to retrieve the composition of the planet's atmosphere with high fidelity. Furthermore, even if the atmospheric composition is known, data interpretation remains problematic and potentially degenerate \citep{Cockell2014, Rein2014}. 

Habitability and the emergence of life are complex and poorly understood phenomenons that are influenced by geophysical, chemical, and biological processes. Our definition of life is descriptive and not hypothesis-driven \citep{Lovelock1965, Margulis1995, McKay2004, Trifonov2012, Bains2014}. That is, we know the attributes of life (e.g., metabolism and adaptation), but we cannot explain it or predict why/how/when it emerges. Our only example of a habitable planet is Earth, and we do not know how chemistry transitioned to biology here. We also do not know whether life exists within `our reach' in the Solar System e.g., on Mars or below the ice of Europa. Thus, we need to be cautious about how we approach the topic of life on exoplanets based on the limited but steadily growing observational data.

The question of habitability is often simplified to the concept of the Habitable Zone (HZ): can the planet harbor liquid water (a crucial ingredient for life as we know it) on its surface? This is still a difficult question to answer. To date, the only observational data that helps to constrain the surface climate are the planet's mass and/or radius, and the stellar properties, most importantly the stellar flux that reaches the planet\footnote{The luminosity and mass of the host star combined with the orbital parameters of the planet yield the incoming stellar flux.}. Although these data are necessary to constrain the surface climate, they alone are insufficient to do so. 

The surface climate of a planet, and consequently its HZ, is strongly influenced by the properties of the atmosphere, geophysics, and the planet formation history. In fact, there are several HZ descriptions because different authors make different assumptions about planet properties. In other words, the HZ is planet-specific. For example, the geophysical process called the carbon-silicate cycle is believed to regulate the atmospheric CO$_2$ budget on Earth \citep{Walker1981}. The earliest work on the HZ \citep{Hart1979} did not include the carbon-silicate cycle and their HZ limits were narrow compared to later work with the carbon-silicate cycle \citep{Kasting1993, Kopparapu2013}. Similarly, the atmosphere has a strong influence on the HZ. A rocky planet with primordial or outgassed H$_2$ can be habitable at large distances from a host star compared to Earth-like planets with N$_2$/CO$_2$/H$_2$O atmospheres \citep{Pierrehumbert2011}. Clouds \citep[see e.g.,][]{Selsis2007, Yang2013} and the water reservoir of the planet \citep[see e.g.,][]{Abe2011, Leconte2013, Zsom2013} also have a strong impact on the HZ . 

As the atmospheric and geophysical properties of potentially rocky exoplanets are currently unknown, it is not possible to decide which HZ description should be favored over others. That is, one data point -- Earth -- does not carry enough information to tell which habitable planet scenario is the most frequent in the Universe \citep[see the argument of][on a similar problem]{Spiegel2012}, and which corresponding HZ limit should be used in e.g., HZ occurrence rate estimates.

The goal of the paper is to develop a population-based HZ description that does not select one specific planet scenario, but instead it considers a wide range of planet scenarios. In that case, the HZ takes the form of a probability function with respect to observationally constrained stellar and planet properties. For example, the model describes the probability that an Earth-like exoplanet has liquid water on its surface if we have no prior knowledge about the planet's atmospheric and geophysical properties. A planet is Earth-like, if its radius is one Earth radius (1 $R_\oplus$), and it receives one solar constant incoming stellar flux (1 $F_\oplus$). Given our current observational knowledge of exoplanets, the constraints on the HZ are weak. However, the advantage of the method described here is its flexibility: when the atmospheric properties of potentially habitable rocky planets become available, the population-based HZ will be a function of more observationally constrained properties, and the constraints on the HZ will be stronger. We illustrate the flexibility of the method through an example in Sec. \ref{sec:discussion}.

The paper is structured as follows. The methods are described in Sec. \ref{sec:methods} with special attention on how the observationally unconstrained planet parameters are treated. The results are discussed in Sec. \ref{sec:results}, and areas of further improvements are explored in Sec. \ref{sec:discussion}. Finally, the results are summarized in Sec. \ref{sec:sum}.

\begin{table*}[h]
\caption{Adapted probability density functions of observationally unconstrained random variables. A synthetic planet population is constructed for over 4000 possible combinations of these PDFs. The algorithm used to synthesize the planet population is depicted in Fig. \ref{fig:flowchart}. Each PDF has a short label (in brackets) which is used in Fig. \ref{fig:impact} where the impact of random variables on the surface climate is assessed.}
\label{tbl:PDFs}
\centering
\begin{tabular}{@{\vrule height 10.5pt depth2pt  width0pt}ll}
\hline
Random variable & PDF \cr
\hline\hline
Planet mass & optimistic with $\sigma = \mathrm{mean}/6$  (PDF Pm1)\cr
 & nominal with $\sigma = \mathrm{mean}/6$ (PDF Pm2)\cr
  & pessimistic with $\sigma = \mathrm{mean}/6$ (PDF Pm3)\cr
  & optimistic with $\sigma = \mathrm{mean}/3$  (PDF Pm4)\cr
 & nominal with $\sigma = \mathrm{mean}/3$ (PDF Pm5)\cr
  & pessimistic with $\sigma = \mathrm{mean}/3$ (PDF Pm6)\cr
\hline
Atmosphere type & H$_2$, N$_2$, CO$_2$ with equal probability (PDF At1)\cr
 & N$_2$, CO$_2$ with equal probability (PDF At2)\cr
 & only N$_2$ (PDF At3)\cr
 & only CO$_2$ (PDF At4)\cr
 & only H$_2$ (PDF At5)\cr
\hline
Surface pressure & uniform in $\log$, P$_{\textrm{min}} = 1$ Pa (PDF Sp1)\cr
 & uniform in $\log$, P$_{\textrm{min}} = 10^2$ Pa (PDF Sp2)\cr
 & uniform in $\log$, P$_{\textrm{min}} = 10^4$ Pa (PDF Sp3)\cr
 & lognormal with mode at 0.1 bar (PDF Sp4)\cr
 & lognormal with mode at 1 bar (PDF Sp5)\cr
 & lognormal with mode at 10 bar (PDF Sp6)\cr
 & lognormal with mode at 100 bar (PDF Sp7)\cr
\hline
Surface albedo & uniform (PDF Sa1)\cr
 & normal with $\bar{a}_{\textrm{surf}}$ = 0.2, $\sigma = 0.1$ (PDF Sa2)\cr
 & lognormal with mode = 0.1 (PDF Sa3)\cr 
\hline
Relative humidity & uniform in $\log$ (PDF Rh1)\cr
 & uniform (PDF Rh2)\cr
 & lognormal with mode at 50\% (PDF Rh3)\cr
\hline
N$_2$ mixing ratio\footnote{These PDFs are sampled if the atmosphere is CO$_2$ dominated. The range of N$_2$ mixing ratio is between 0 or 10$^{-5}$ and 0.5, CO$_2$ comprises the rest of the atmosphere.} & uniform (PDF mr1)\cr
 & uniform in $\log$ (PDF mr2)\cr
 & lognormal with mode at $4\times 10^{-2}$ \footnote{CO$_2$ mixing ratio on Earth.} (PDF mr3)\cr
\hline
CO$_2$ mixing ratio\footnote{These PDFs are sampled if the atmosphere is N$_2$ dominated. The range of CO$_2$ mixing ratio is between 0 or 10$^{-5}$ and 0.5, N$_2$ comprises the rest of the atmosphere.} & uniform (PDF mr1)\cr
 & uniform in $\log$ (PDF mr2)\cr
 & lognormal with mode at $3.5 \times 10^{-2}$ \footnote{N$_2$ mixing ratio on Venus.} (PDF mr3)\cr
\end{tabular}
\end{table*}

\begin{figure*}[h]
\centering
\caption{The flowchart of HUNTER. The goal of the algorithm is to synthesize planetary and atmosphere populations using a set of random variables and their probability density functions. The algorithm distinguishes Super Earth planets from rocky planets, checks whether the planet's surface pressure is not too small or not too large, and finally, if all requirements are fulfilled, the surface climate is estimated. All samples are saved and the data is evaluated to assess the Habitable Zone.}
  \includegraphics[width=0.7\textwidth]{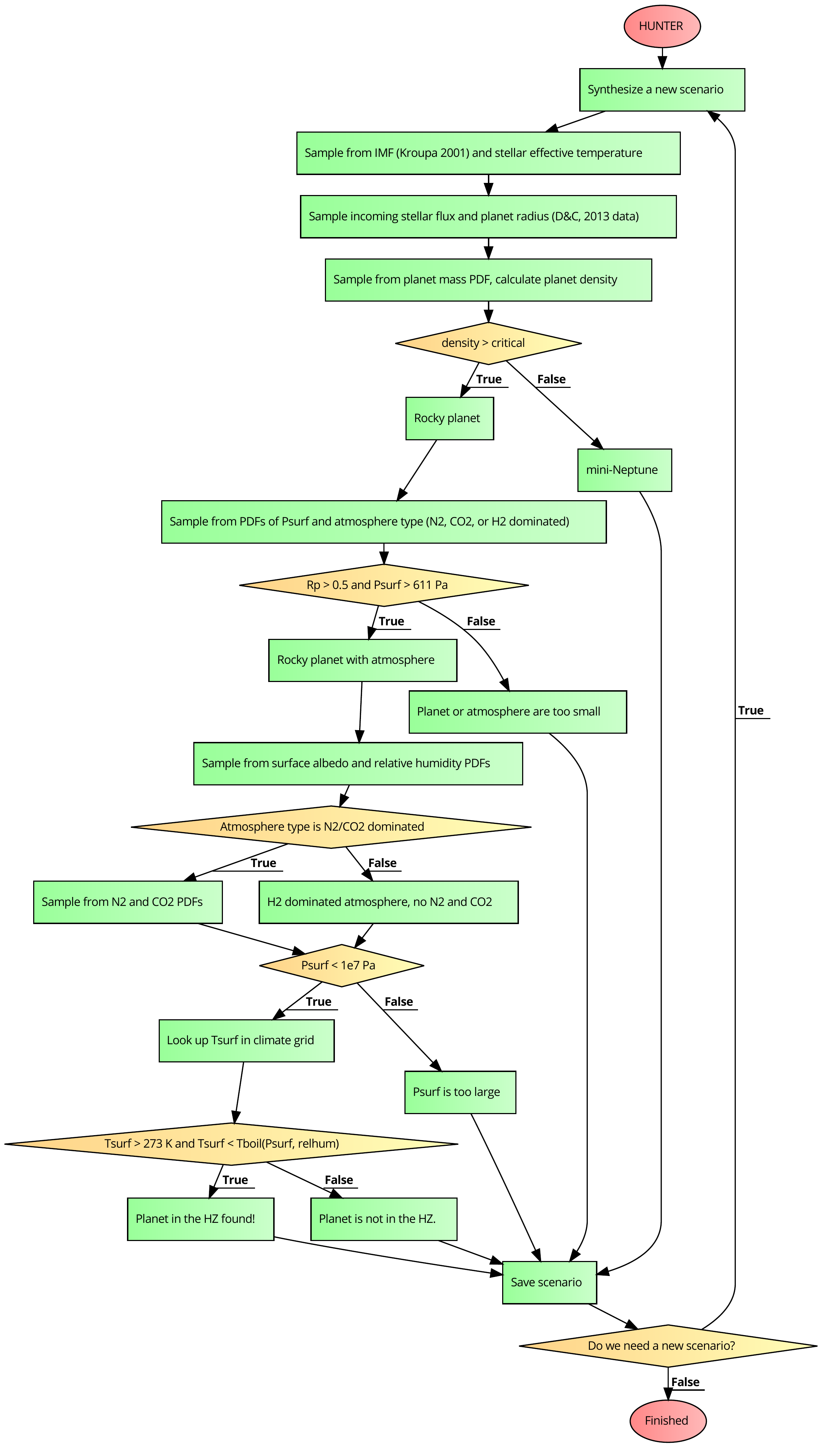}
\label{fig:flowchart}
\end{figure*}

\section{Methods}
\label{sec:methods}
Whether a planet has liquid water on its surface depends on its surface pressure and temperature, and the surface temperature in turn is influenced by a large number of other parameters. If the surface pressure is too low or too high, water cannot exist in liquid form based on the water phase diagram. The surface temperature itself depends on the incoming stellar flux (a function of stellar type and semi-major axis), the surface pressure, the atmospheric composition, surface gravity (function of planet mass and radius), and surface albedo. The latter four properties describe the greenhouse effect of the atmosphere within the limit of a 1D vertical climate model\footnote{Additional properties that impact the surface climate are e.g., the rotation period and the obliquity of the planet, which are important in 3D global circulation models.}, i.e., how much larger the surface temperature is compared to the equilibrium temperature of the planet\footnote{The equilibrium temperature of Earth and Venus are similar (252 K and 240 K, respectively). However, the surface temperatures radically differ (290 K for Earth and in excess of 700 K for Venus) because the atmosphere of Venus has a much stronger greenhouse effect than Earth's atmosphere.}. 

The main concept of the method is to treat each parameter that influences the surface climate as a random variable with assigned probability density functions (PDFs). The PDF describes the relative likelihood that a random variable takes on a given value. Some variables (such as the stellar properties, planet radius, and incoming stellar flux) are constrained by observations. Other parameters (such as the surface pressure, relative humidity) are observationally unconstrained. 

Then, the population-based HZ describes the probability ($p_{\mathrm{HZ}}$) that an exoplanet harbors liquid water as a function of the observationally constrained variables. Currently, these are the stellar type ($ST$), planet radius ($R_p$), and the incoming stellar flux ($F_{\mathrm{in}}$), thus $p_{\mathrm{HZ}} = p_{\mathrm{HZ}}(ST, R_p, F_{\mathrm{in}})$ and no prior assumptions on the exoplanet's atmosphere and geophysical properties are made. The focus is on transiting planets in this paper, that is why the planet radius is treated as an observationally constrained variable and not the planet mass. Although the radial velocity technique can measure the masses of small planets \citep{Bonfils2013a}, the sample of potentially habitable transiting planets is larger and better suited for a statistical analysis.

The procedure to estimate $p_{\mathrm{HZ}}$ is summarized in this paragraph, and a more detailed description is given in the rest of this section. $p_{\mathrm{HZ}}$ and its uncertainty are estimated in the following way. A large number of potential PDFs are adapted for each observationally unconstrained random variable (see Sect. \ref{sec:PDFs} and Table \ref{tbl:PDFs} for more details). The observationally unconstrained random variables are 1) the planet mass, 2) atmosphere type, 3), surface pressure, 4) surface albedo, 5) relative humidity, 6) N$_2$ or CO$_2$ mixing ratios, if the atmosphere is CO$_2$ or N$_2$ dominated, respectively. The PDFs reflect our current knowledge and expectations about exoplanet properties. For example, some PDFs describe silicate-iron planet cores, water worlds, planets with hydrogen atmospheres, planets with Earth-like atmospheres, etc. One PDF is selected for each random variable, and we loop through all possible combinations of PDFs. There are over 4000 combinations. For each PDF combination, a synthetic exoplanet population is created with typically 10$^5$ exoplanet scenarios per population. The flow chart of the algorithm is illustrated in Fig. \ref{fig:flowchart} and explained in more detail in Sec. \ref{sec:PDFs}. The goal of the algorithm is to distinguish predominantly gaseous mini-Neptunes from rocky planets, identify planets that have too small or too high surface pressures. Finally, a 1D vertical climate model is used to calculate the surface temperature of planet scenarios with the potential to harbor liquid water (see Sect. \ref{sec:climate}). Then, the subpopulation of exoplanets that can harbor liquid water on their surfaces is analyzed. A 2D $F_{\mathrm{in}}$-$R_p$ grid is generated for each stellar type considered, and we count what fraction of exoplanets are habitable within each grid cell. As there are over 4000 populations, over 4000 different estimates on $p_{\mathrm{HZ}}$ are generated, which allows us to constrain the uncertainty of our results: the distribution of $p_{\mathrm{HZ}}$ within each $F_{\mathrm{in}}$-$R_p$ grid cell for each stellar type.

The python implementation of the method is called HUNTER (Habitable zone UNcerTainty EstimatoR) and it is publicly available\footnote{https://github.com/andraszsom/HUNTER}. HUNTER is designed in such a way that it is easy to implement new observational results or to experiment with new types of PDFs.

\subsection{Probability Density Functions of Random Variables}
\label{sec:PDFs}

\paragraph{Sampling incoming flux and planet radius}
We use two different methods to sample the incoming stellar flux and the planet radius. 1) The incoming stellar flux and planet radius of the exoplanet population are drawn from distributions that are uniform $\log F_{\mathrm{in}}$ and $\log R_p$. These two distributions are best suited to study the general shape and properties of $p_{\mathrm{HZ}}$. If $F_{\mathrm{in}}$ and $R_p$ are uniform in log space, it is possible to study $p_{\mathrm{HZ}}$ at regions of the parameter space where exoplanets have not been discovered yet (e.g., exoplanets with $F_{\mathrm{in}} < 0.1 F_{\oplus}$). 2) $F_{\mathrm{in}}$ and $R_p$ are sampled from observed distributions. The second method is used to study the occurrence rate of potentially habitable planets. We use the \cite{Dressing2013} data set of transiting planets orbiting M dwarfs, rather than similar data sets for sunlike stars \citep{Petigura2013, Foreman-Mackey2014}. The former sample is more complete to planets with cooler temperatures than Earth, and does not rely on interpolation in this temperature range to the same extent as for G dwarfs. This is dually true because both the geometric transit likelihood and the signal-to-noise per transit increase for cool planets orbiting M dwarfs, as opposed to G dwarfs.

\begin{figure}
\centering
\includegraphics[width=0.49\textwidth]{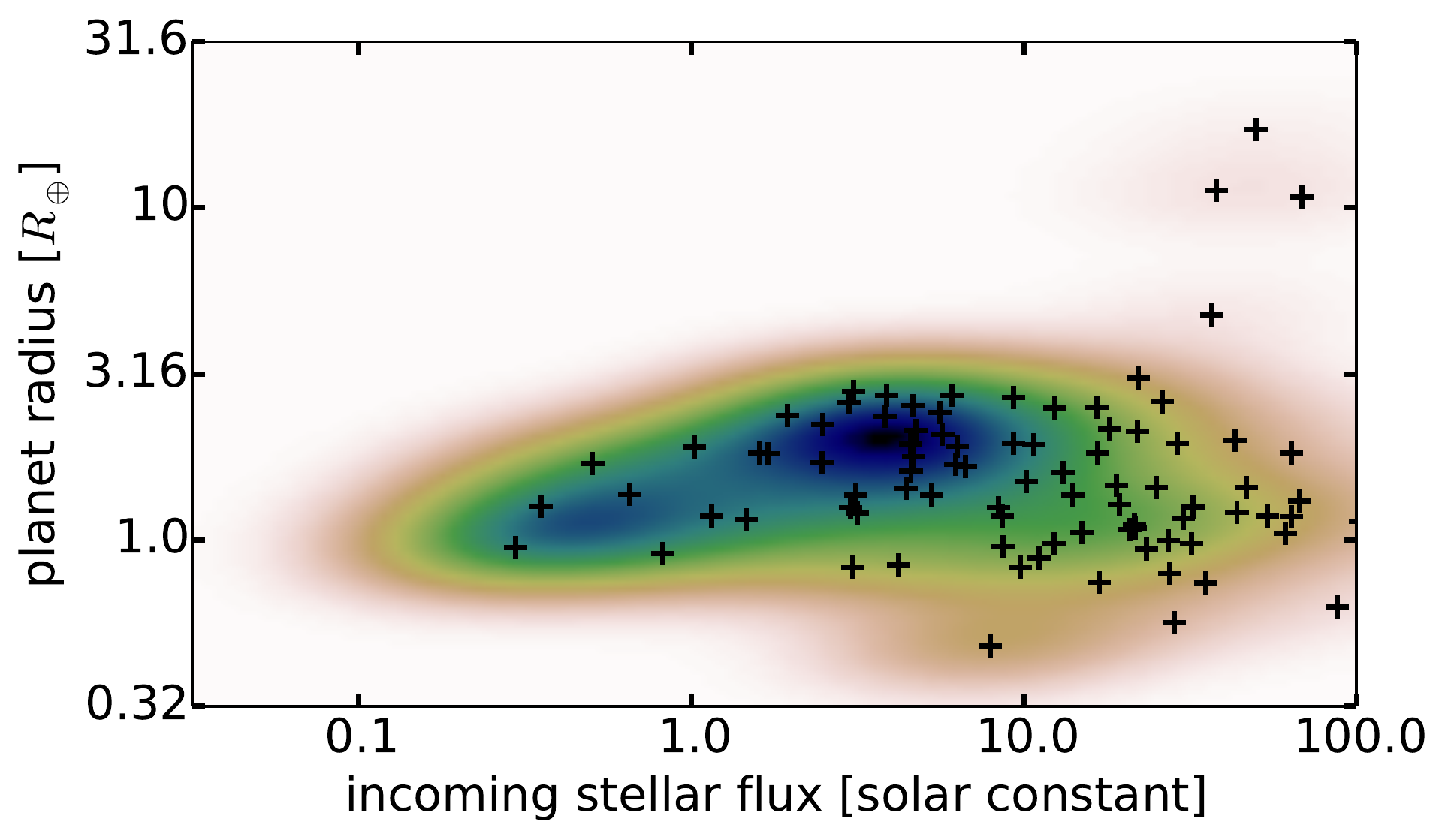}
\caption{The joint probability density function of incoming stellar flux and planet radius (contour levels), and the planet radius and incoming stellar flux data of \cite{Dressing2013} (`+' signs). Each data point is weighted with the inverse of the transit probability to correct for the observational bias.}\label{fig:KDE}
\end{figure}

The data of \cite{Dressing2013} is recast using Gaussian kernel density estimation. This was necessary for two reasons. The PDF (or occurrence rate) is given as a function of orbital period and planet radius in \cite{Dressing2013}. However, occurrence rates as a function of incoming stellar flux and planet radius are more useful for climate modeling. The PDF values were described on a coarse grid, which is not well-suited for random sampling. For these reasons, a Gaussian kernel density estimator is used to characterize the smooth probability density as a function of the incoming stellar flux and planet radius. The joint PDF is illustrated in Fig. \ref{fig:KDE}. 

\begin{figure}
\centering
\includegraphics[width=0.49\textwidth]{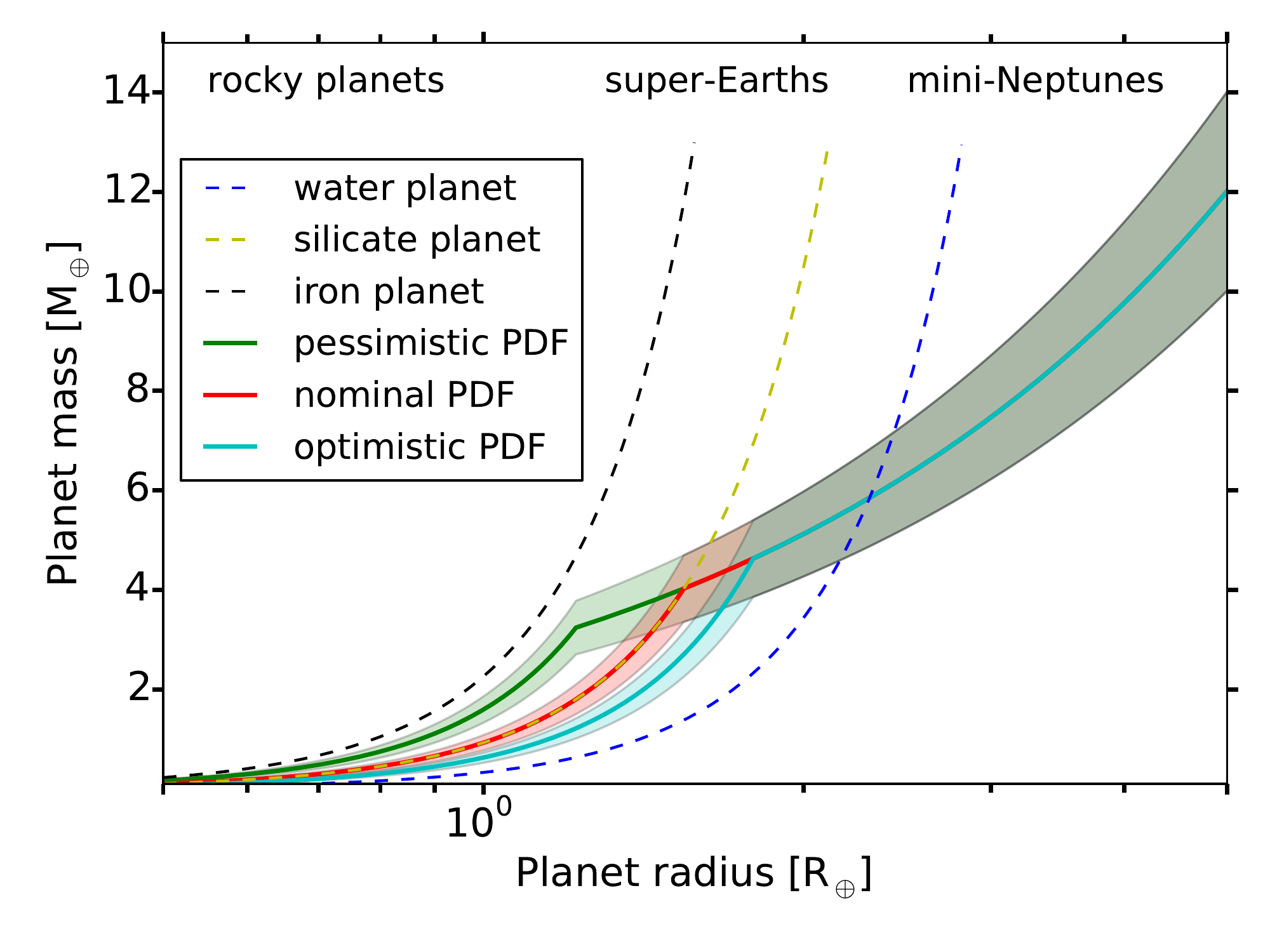}
\caption{The radius-mass diagram of three planet mass probability density functions with $\sigma = \mathrm{mean}/6$ (PM1, PM2, PM3 distributions from Table \ref{tbl:PDFs}). The planet radius is an observationally constrained parameter from e.g., the Kepler data of \cite{Dressing2013}. Thus, we are interested in the planet mass PDF as a function of planet radius. Each PDF follows a Gaussian distribution where the mean is depicted by the solid lines and the standard deviation is illustrated by the shaded region. Each PDF has a small and large planet region. In the small planet region (left from the `kink' in the curve), all planets are rocky. Boundaries are imposed such that no planet can be denser than pure iron or less dense than pure water in this region. If the planet radius is larger than a critical value (right from the `kink' in the curve), a planet can be a super-Earth or a mini-Neptune depending on its density. Planets in this region cannot be denser than iron. The radius-mass relationships of pure iron, silicate, and water worlds are also plotted as reference (dashed lines).}
\label{fig:pmPDF}
\end{figure}

\paragraph{The probability density function of planet mass} The planet radius is an observationally constrained parameter, but the planet mass is unknown for most small exoplanets at habitable distances from the star. Therefore, planet mass PDFs are used to sample the planet mass based on the planet radius. Some planet mass PDFs are illustrated in Fig. \ref{fig:pmPDF}. All PDFs follow a Gaussian distribution around some mean with a standard deviation which is a fraction of the mean. Small rocky planets are distinguished from large planets that can be super-Earths or mini-Neptunes depending on their bulk density. The PDFs can be optimistic, nominal, and pessimistic. The PDFs are optimistic or pessimistic with respect to the occurrence rate of water worlds. The properties of the PDFs are described below. To guide how the PDFs are created, the mass of pure iron, pure silicate, and pure water planets are illustrated as a function of radius R$_p$ (M$_i(R_p)$, M$_s(R_p)$, M$_w(R_p)$, respectively) following \cite{Seager2007}. 

In the optimistic distribution, the planet mass follows a Gaussian distribution if the planet radius is below 1.79 R$_\oplus$. The mean of the distribution is $\bar{M}_p = (M_s(R_p) + M_w(R_p))/2$, the standard deviation can be $\sigma = \bar{M}_p / 6$ or 3. M$_p$ is not allowed to be smaller than M$_w(R_p)$ or larger than M$_i(R_p)$. Effectively, such a description means that all planets below 1.79 $R_\oplus$ have a surface (either solid or liquid) and the occurrence rate of pure water planets is significant. If the planet radius is larger than 1.79 $R_\oplus$, the mean of the distribution is prescribed by an empirical mass-radius relationship (Eq. 3 in \cite{Weiss2014}): $\bar{M}_p = 2.69 R_p^{0.93}$, where $M_p$ and $R_p$ are expressed in Earth mass and radius units, respectively. The standard deviation is again $\sigma = \bar{M}_p / 6$ or 3. Planets are not allowed to be denser than iron. However, the lower mass boundary is given by M$_w(1.79 R_\oplus)$. The critical density is given by the density of a pure water planet and all planets denser than the critical density are considered rocky. 

It might appear that the optimistic distribution contradicts Solar System observations because the inner planets are rocky. However, some moons of Saturn (e.g., Mimas and Thetys) are primarily made out of water ice and only small amounts of rock (the density of Thetys is less than that of water). Although these are small moons, they indicate that water-rich worlds could form at larger distances from the star. If such planets have hydrogen-dominated atmospheres, they could harbor liquid water on their surfaces. 


The nominal and pessimistic mass distributions differ in three aspects. In the nominal distribution, if the planet radius is below 1.54 $R_\oplus$, $\bar{M}_p = M_s(R_p)$, and planets are not allowed to be smaller than $(M_w(R_p)+M_s(R_p))/2$. In other words, there are no pure water planets, but there exists half water - half silicate rocky planets. The third difference is that the critical density is given by the density of a half water - half silicate planet. In the pessimistic distribution, $\bar{M}_p = (M_i(R_p) + M_s(R_p))/2$ below 1.22 $R_\oplus$, and planets are not allowed to be smaller than M$_s(R_p)$. That is, most rocky planets are assumed to be similar to the rocky planets in our Solar System (without a significant water layer). The critical density in this case is given by the density of a pure silicate planet.

Recently, \cite{Rogers2015} showed that most 1.6 $R_\oplus$ planets are not expected to be rocky (silicate-iron composition), which at first glance would indicate that the optimistic distribution (with all planets below 1.8 $R_\oplus$ considered to have a surface) is inconsistent with observations. It is important to note that planets drawn from the optimistic distribution are expected to be volatile-rich (not silicate and iron-rich), thus there is no contradiction here. Furthermore, all exoplanets used in the work of \cite{Rogers2015} receive more than 1.1 $F_\oplus$. It is likely that more distant exoplanets (with $F_{\oplus} < 1$) have different properties because such planets might have formed outside the snow line and thus are more volatile rich (similarly to the large moons of Jupiter and Saturn). For these reasons, we believe that the optimistic distribution is not inconsistent with observations.

\paragraph{Atmosphere type} If the planet is rocky (its density is larger than the critical density prescribed by the planet mass PDF), the atmosphere type and surface pressure of the planet are sampled. Five different PDFs are considered for the atmosphere type: H$_2$, N$_2$, CO$_2$ dominated atmospheres are equally likely; there are no H$_2$ dominated atmospheres and N$_2$ or CO$_2$ dominated atmospheres are equally likely; there are only N$_2$ dominated atmospheres; only CO$_2$ dominated atmospheres; or only H$_2$ dominated atmospheres. Some of these PDFs represent extremes. For example, it would be surprising if N$_2$ or CO$_2$ dominated atmospheres in the HZ are the exception and H$_2$ dominated atmospheres are frequent. However, such a PDF cannot be excluded based on our current knowledge. 

\paragraph{Surface pressure} Next, a sample is drawn from a surface pressure distribution and it is important to consider what the range of surface pressures should be. The minimum surface pressure under which water could exist in liquid form is set by the triple point pressure of water (611 Pa). If the planet's surface pressure is less than 611 Pa, water exists only either in ice or vapor form. Some PDFs are uniform in $\log(P)$ (see Table \ref{tbl:PDFs}). In this case, the minimum surface pressure is a parameter and PDFs with $P_{\mathrm{min}}$ = 1, 10$^2$, 10$^4$ Pa are considered. The surface pressure on Mars is 600 Pa for comparison. The minimum surface pressure is set to 0 Pa for lognormal PDFs with modes (most frequently occurring values) of 10$^4$, 10$^5$, 10$^6$, 10$^7$ Pa. 

The maximum surface pressure is set by a condition on the total atmospheric mass. The correlation between the atmosphere mass ($m_{\mathrm{atm}}$) and surface pressure ($P_{\mathrm{surf}}$) is expressed as
\begin{equation}
P_{\mathrm{surf}} = \frac{m_{\mathrm{atm}}g_{\mathrm{surf}}}{4\pi R_p^2},
\end{equation}
where $g_{\mathrm{surf}}$ is the surface gravity of the planet. If the atmosphere is N$_2$ or CO$_2$ dominated, the maximum surface pressure is such that the atmosphere mass is 1\% of the planet's mass. If the atmosphere is H$_2$ dominated, the maximum surface pressure is set by $m_{\mathrm{atm}} = 10^{-5} M_p$.  

Examples are shown to validate the choices of maximum atmosphere masses and surface pressures. If the atmosphere mass is 1\% on an Earth-like planet, the surface pressure is $10^4$ bars. The effective height\footnote{The effective height describes a circle with radius $r_p + \Delta r (\lambda)$ (where $r_p$ is the planet radius and $r_p + \Delta r (\lambda)$ is the wavelength dependent radius of the circle) that effectively blocks as much light from the stellar disk as the planet's atmosphere does.} of such an atmosphere is only 100 km higher than the effective height on Earth (assuming a constant 10 km pressure scale height). Therefore, the relative difference in radius between an Earth-like planet with $P_{\mathrm{surf}}$ = 1 bar and $P_{\mathrm{surf}}$ = 10$^4$ bars is only $\sim 2$\%. The typical uncertainty in the measured planet radius is on the order of 10\% in the data of \cite{Dressing2013} mostly due to uncertainty in the stellar radius. Thus, currently we are unable to distinguish an Earth-analog planet from an Earth-like planet with a surface pressure of 10$^4$ bars around M dwarfs. The situation is different, if Earth hypothetically had an H$_2$ dominated atmosphere. If $m_{\mathrm{atm}} = 10^{-5} M_\oplus$, the surface pressure is 10 bars. The pressure scale height of H$_2$ atmospheres is an order of magnitude larger than N$_2$ atmospheres of the same temperature due to the low mean molecular weight of H$_2$. If Earth had a 10 bar H$_2$ atmosphere, the atmosphere's effective height would be $\sim$ 200 km. Although the relative radius difference is larger than in the previous example ($\sim 4$\%), it is still below the typical uncertainty of \cite{Dressing2013}.

\paragraph{Surface albedo and relative humidity} If the radius of the rocky planet is larger than 0.5 $R_\oplus$ and the surface pressure is above the critical point pressure of water, then we proceed to the next step of our habitability investigation. The lower radius limit is set to avoid sampling exoplanets that have low or negative radii: the kernel density estimator can extrapolate to negative radii exoplanets. Therefore, the lower radius limit is set by the exoplanet with the smallest radius in the data set of \cite{Dressing2013}. In the next step, the surface albedo and relative humidity PDFs are sampled. Three surface albedo PDFs are used: uniform, normal, and lognormal (see Table \ref{tbl:PDFs} for more details). The surface albedo is between 0 and 0.5 around M dwarfs, and between 0 and 1 around Solar-like stars. The albedo upper limit around M dwarfs is set by the reduced albedo of ice and snow at near infrared wavelengths where the spectral energy distribution of M dwarfs peak \citep{Joshi2012, Shields2013}. 

The relative humidity PDF can be uniform between 1\% and 100\%, uniform in $\log$ space, and lognormal with a mode at 50\%. A relative humidity value of 50\% is the effective relative humidity in the troposphere of Earth. That is, the global average vertical relative humidity profile of Earth \citep{Manabe1967} can be approximated with a constant effective relative humidity of 50\% to reproduce the global average surface temperature of Earth \citep{Pierrehumbert2010}.

\paragraph{N$_2$ and CO$_2$ mixing ratios} If the atmosphere is N$_2$ or CO$_2$ dominated, the mixing ratios of these two components are sampled. In an N$_2$ dominated atmosphere, the mixing ratio of CO$_2$ is between 0 and 0.5. The PDF of CO$_2$ mixing ratio can be uniform, uniform in $\log$ space, and lognormal with the mode of $4\times10^{-2}$, which is the CO$_2$ mixing ratio on Earth. Similarly, if the atmosphere is CO$_2$ dominated, the mixing ratio of N$_2$ is between 0 and 0.5, and similar PDFs are adapted as before with the difference that the mode of N$_2$ in the lognormal distribution is $3.5\times10^{-2}$, which is the N$_2$ mixing ratio in the atmosphere of Venus.

\paragraph{Surface temperature} If the surface pressure is below $10^7$ Pa, we investigate whether the surface temperature is suitable for liquid water. In principle, water can be liquid at pressures above $10^7$ Pa. However, the plausibility to detect biosignature gases in dense atmospheres decreases with surface pressure \citep{Seager2013a}. The reason is that more biomass is needed to produce the same biosignature mixing ratio, if the surface pressure is large. If Earth had a surface pressure of $10^7$ Pa, roughly 100 times more biomass would be necessary to reproduce the current atmospheric O$_2$ and CH$_4$ mixing ratios, which is unlikely to occur. 

The surface temperature is not drawn from a PDF but it is calculated from all the parameters sampled up to this point. The reason is that the sampled values unequivocally determine the surface temperature within the limits of a 1D vertical climate model. The incoming stellar flux in combination with the stellar, planetary, and atmosphere parameters allow us to estimate the surface temperature. The method is described in the next Section.

Finally, we note that as the surface pressure cutoff for the atmosphere mass and for the surface temperature check are motivated by observability, $p_{\mathrm{HZ}}$ and the HZ occurrence rates should be considered as observable quantities. In principle, water can exist in liquid form at surface pressures higher than 100 bars. But it is potentially very challenging to identify liquid water on the surfaces of such exoplanets.

\subsection{Climate modeling}
\label{sec:climate}
HUNTER does not perform climate calculations on the fly. Instead, a large set of atmosphere models are pre-calculated in the relevant range of parameter space and store the results in look-up tables that are interpolated in run-time. Such a procedure saves computational time. The look-up tables need to contain the surface temperature as a function of incoming stellar flux and other parameters. The grid dimensions are atmosphere type (three types of atmospheres), surface pressure (between 611 Pa and 10$^7$ Pa covered by 5 grid points), surface gravity (between 2 m/s$^2$ and 45 m/s$^2$, 6 grid points), relative humidity (between 1\% and 100\%, 10 grid points), surface albedo (5 grid points), N$_2$ and CO$_2$ mixing ratios (between 0 and 0.5, 6 grid points), and surface temperature (between 273 K and the boiling point, which is surface pressure and relative humidity dependent, the grid has a step size of 1 K). The code loops through these grids and generate all-convective temperature-pressure profiles using the climate model of \cite{Zsom2013}. The climate model calculates the incoming stellar flux under which the atmospheres are in radiative-convective equilibrium and the flux values are stored in a look-up table for HUNTER. The assumption of an all-convective profile is valid if one is interested in the surface climate only. The reason is that the other parts of the atmosphere are largely optically thin and have a small effect on the surface temperature \citep{Pierrehumbert2011}. A modified version of the 1D climate model described in \cite{Zsom2013} is used to generate the look-up tables. It is a globally averaged vertical climate model that includes greenhouse gas absorption, Rayleigh scattering, and collision-induced absorption of N$_2$, CO$_2$ and H$_2$. 

HUNTER interpolates the look-up tables to determine the surface temperature at each sampled scenario. If the surface temperature is between 273 K and the boiling point of water, the planet is habitable. Otherwise, the surface temperature is either too large or too small and water is in vapor or ice form, respectively. As the surface temperature in the tables is restricted to the freezing and boiling points of water, the surface temperature of planets that cannot harbor liquid water is not calculated. 

\subsection{Atmospheric stability}
We perform an analysis on atmosphere stability for exoplanets orbiting M dwarfs. We follow the description of \cite{Heng2012} to evaluate whether exoplanet atmospheres in the synthetic population are stable against atmospheric collapse. We assume that the planets are synchronously rotating and we calculate the radiative and advective time scales of the atmosphere. If the advective time scale is shorter than the radiative timescale, the atmosphere is considered stable. \cite{Heng2012} use the irradiation temperature and the photospheric pressure to calculate the radiative time scale, we use the surface temperature and the surface pressure. The advective time scale is estimated to be $t_\mathrm{conv} = R_p / c_s$, where $c_s$ is the sound speed at the surface of the exoplanet. In general, we find that atmospheres with low surface pressure and high mean molecular weight are unstable in agreement with \cite{Heng2012}. We do not study the stability of exoplanet atmospheres around Sun-like stars because their rotation period do not correlate with their orbital period.

\section{Results}
\label{sec:results}

\begin{figure*}
\centering
\includegraphics[width=0.49\textwidth]{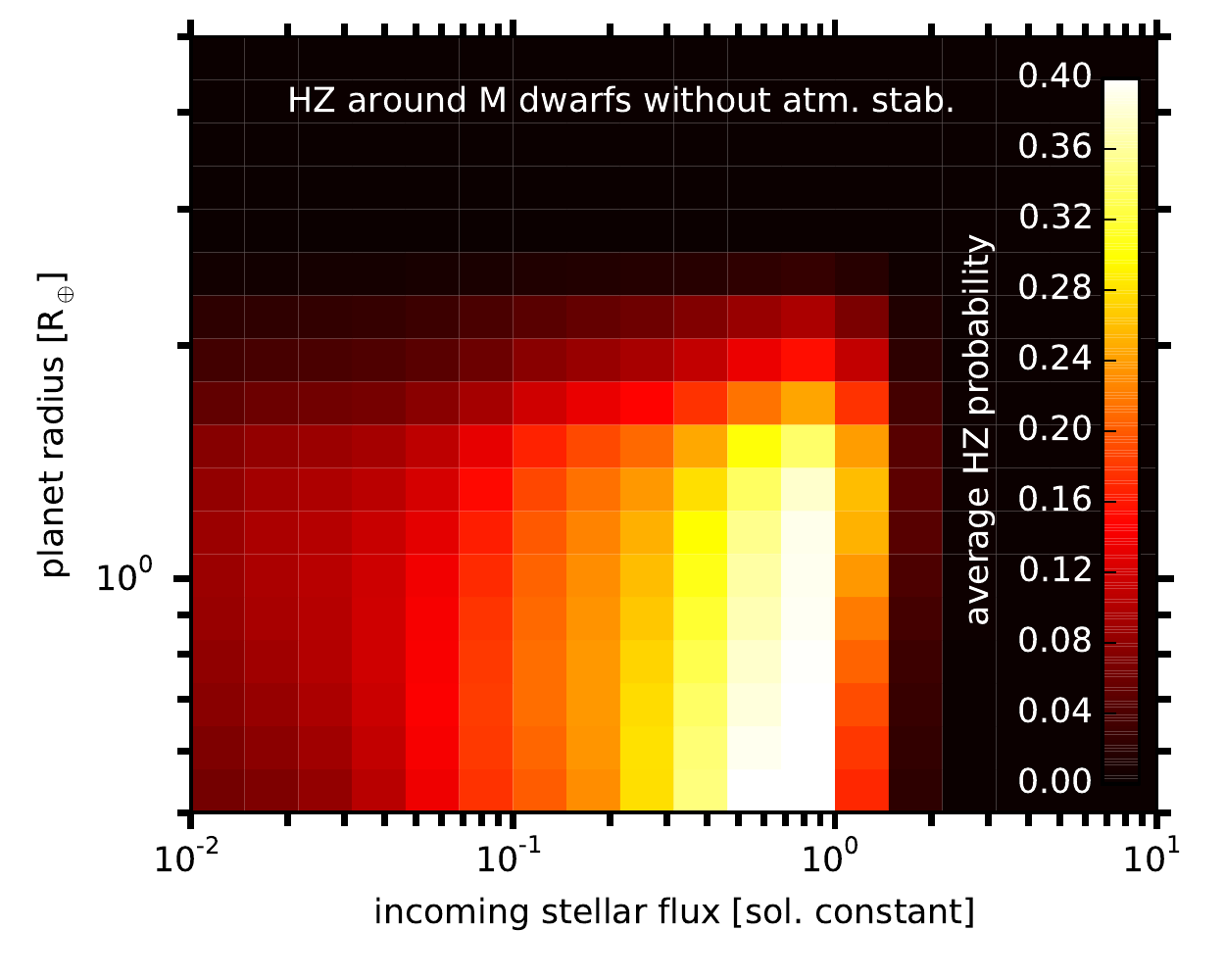}
\includegraphics[width=0.49\textwidth]{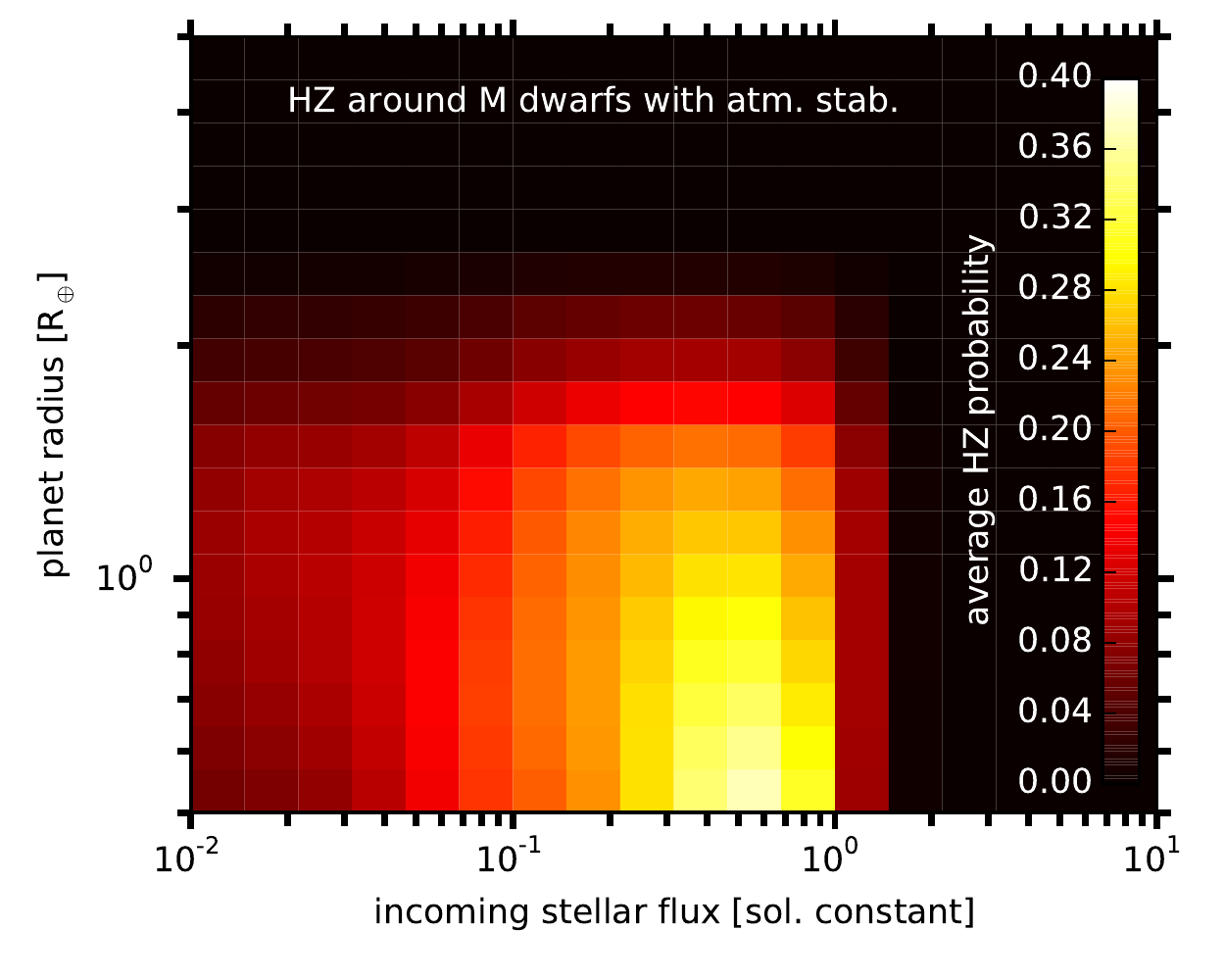}
\caption{The average HZ probability as a function of incoming stellar flux (x axis) and planet radius (y axis) with and without atmospheric stability analysis (left and right panels, respectively). The contour levels correspond to the average HZ probability as indicated on the color bars. The results show that the HZ has a sharp inner edge, but the outer edge can potentially be smooth. The atmospheric stability analysis rules out close-in exoplanets that have low surface pressures and high mean molecular weights. Distant exoplanet with large surface pressures are generally stable agains atmospheric collapse.}\label{fig:HZ}
\end{figure*}

\begin{figure}
\centering
\includegraphics[width=0.49\textwidth]{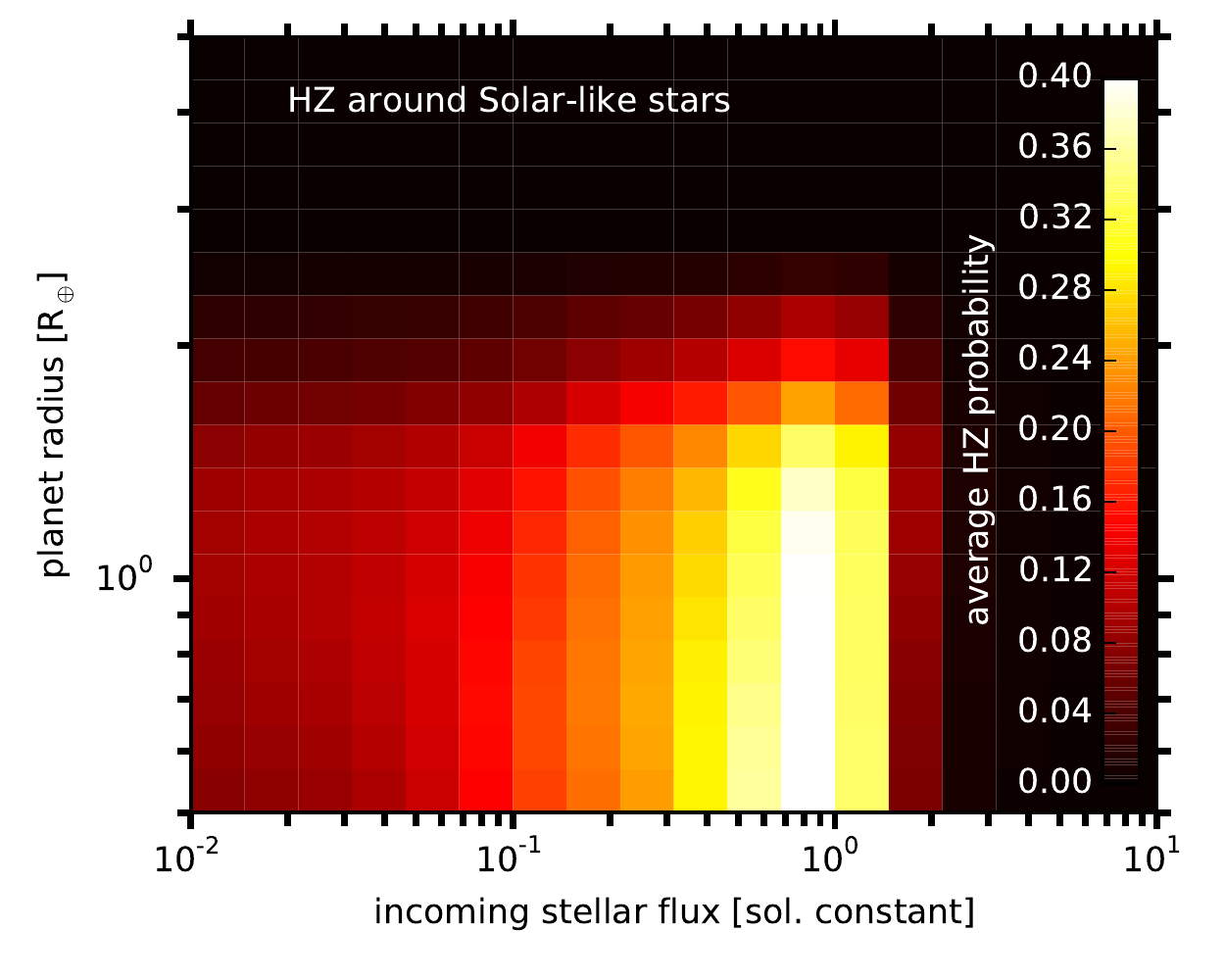}
\caption{The average HZ probability of exoplanets around solar-like stars. The axes and contour levels are the same as in Fig. \ref{fig:HZ}. The figure shows that if the HZ limit is expressed in units of incoming stellar flux, variations in stellar type have a small influence compared to the uncertainty of unknown atmospheric properties. I.e., Fig. \ref{fig:HZ} left and this figure are similar and the main difference is that the HZ inner edge is somewhat closer to the host star for solar-like stars.}\label{fig:HZ2}
\end{figure}


\subsection{Sharp inner edge, potentially smooth outer edge}
\label{subsec:HZ}
The shape of the average HZ probability as a function of incoming stellar flux and planet radius is illustrated in Figs. \ref{fig:HZ} and \ref{fig:HZ2}. The PDFs of planet radius and incoming stellar flux are uniform in log space. Each grid cell contains an average of $\sim 10^6$ synthetic planet scenario and only a fraction of those can harbor liquid water on their surfaces. The shape of $p_{\mathrm{HZ}}$ is calculated for planets orbiting an M dwarf with an effective temperature of 3500 K (Fig. \ref{fig:HZ}, left) and a Solar-like star with an effective temperature of 5900 K (Fig. \ref{fig:HZ}, right). The other difference between the two simulations is the range of surface albedos. The maximum surface albedo in all PDFs is 0.5 for planets around M dwarfs due to the decreased albedo of snow and ice in the near IR. The maximum surface albedo is 1.0 for planets around Solar-like stars. 

There are certain regions of parameter space where it is unlikely that planets harbor liquid water because $p_{\mathrm{HZ}}$ is zero in all scenarios considered: if a planet is larger than 3 $R_{\oplus}$ and if a planet receives more than 3 $F_\oplus$. In the former case, the planet is expected to be a predominantly gaseous mini-Neptune with no solid surface. In the latter case, the planet is strongly irradiated and the surface temperature is above the boiling point of water even if the surface albedo is close to the maximum. In other words, the Habitable Zone has a sharp inner edge. 

Atmospheric stability criteria shows that the atmospheres of some exoplanets orbiting M dwarfs are expected to collapse. The average $p_{\mathrm{HZ}}$ around M dwarfs with and without atmospheric stability analysis is illustrated in Fig. \ref{fig:HZ}. Mostly close-in exoplanets with low surface pressures and/or high mean molecular weights are expected to collapse. Distant exoplanets that have high surface pressures with strong a greenhouse effect are generally stable against collapse.

Potentially habitable inner edge planets orbiting M dwarfs are ideal for atmosphere characterization with transmission spectroscopy for several reasons. The planet to star area ratio is $\sim 1\%$ \citep{Kaltenegger2009}, thus characterization is within reach with the James Webb Space Telescope. The transit probability is high because the planets are close to the star. Finally, such planets have a short orbital period thus transit frequently.

The outer edge of the HZ is potentially smooth because one can find a realistic habitable planet scenario even if the incoming stellar flux is $10^{-2}$ $F_\oplus$. Atmospheres that have high surface pressures ($P_{\mathrm{surf}} > 10$ bars) and high greenhouse gas concentrations can harbor liquid water on their surfaces if the stellar irradiation is small. The atmospheres of such distant exoplanets can be best characterized by direct imaging in emission due to their distance from the star. 

The uncertainty of the HZ probability is large in Figs. \ref{fig:HZ} and \ref{fig:HZ2} (only the average is shown in the Figure). For example, the average probability that an Earth-sized planet that receives 1 $F_\oplus$ is 0.2 and 0.4 around M dwarfs and solar-like stars, respectively. However, the potential range of $p_{\mathrm{HZ}}$ is between 0 and 0.7 for these values. The optimistic estimate of 0.7 is uncertain as well due to the limitations of 1D climate modeling (see Sect. \ref{sec:1Dvs3D} for more details).

It is interesting to consider the atmospheric properties of planets that are Earth-like (Earth-sized planets that receive 1 $F_\oplus$) but not habitable. Such planets can be either too hot or too cold to harbor liquid water. If an Earth-like planet has a thick atmosphere dominated by greenhouse gases such as CO$_2$ or H$_2$, then the surface temperature is above the boiling point of water. On the other hand, if the atmosphere is dominated by gases that are spectroscopically largely inactive (e.g., N$_2$) and the atmosphere contains small amounts of greenhouse gases, then the planet is too cold for liquid water \citep{Zsom2013}.

It is notable how similar Figs. \ref{fig:HZ} left and right are. If the HZ limits are expressed in units of incoming stellar flux, variations in the spectral energy distribution of the star and the surface albedo range have a small impact on the surface climate given that the atmospheric properties of exoplanets are currently unknown. In other words, the impact of the stellar energy distribution on the surface climate becomes important only if the exoplanet's atmosphere and geophysical properties are otherwise well-known. The slight difference is that the inner edge of the HZ is somewhat closer to Solar-like stars than to M dwarfs.

\subsection{Occurrence rate of HZ planets: optimistic upper limits}
\label{sec:occ_rate}

\begin{figure}
\centering
\includegraphics[width=0.49\textwidth]{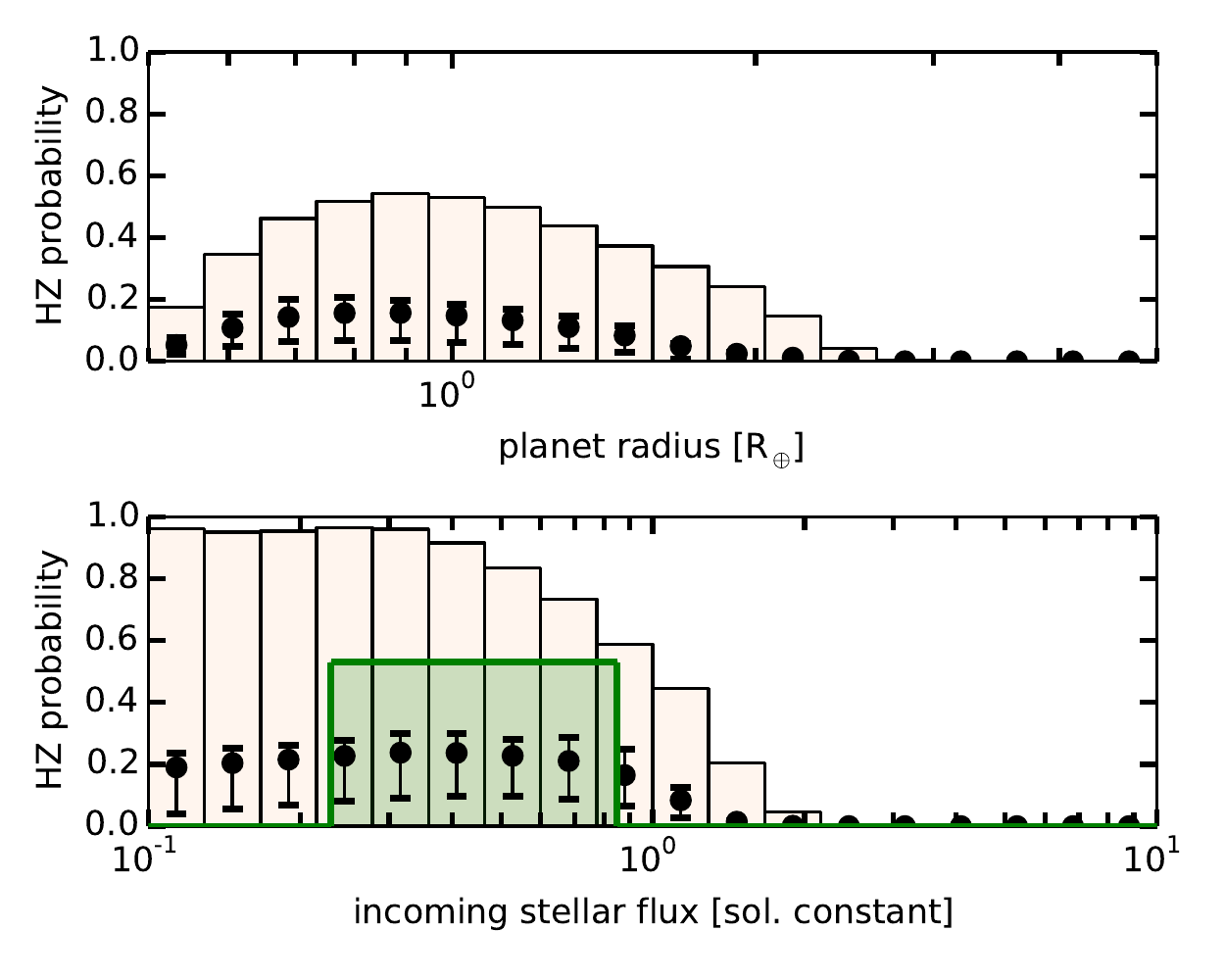}
\caption{The occurrence rate of HZ planets is calculated by integrating the HZ probability ($p_{\mathrm{HZ}}$) over the M dwarfs sample of \cite{Dressing2013}. The figures show the marginalized distributions of $p_{\mathrm{HZ}}$. The top panel shows the fraction of HZ planets marginalized over planet radius. The black dots are the average values. The error bars depict 25 and 75 percentiles, the shaded region depicts the minimum and maximum fractions. The lower panel shows the fraction of HZ planets marginalized over the incoming stellar flux. The symbols are the same as on the top panel. The green shaded region indicates the HZ fraction according to the Earth-analog HZ and assuming that planets between 0.5 and 1.4 $R_\oplus$ are habitable \citep{Kopparapu2013a}. The HZ is a step function in this case. The population-based approach illustrates that there are planets outside the Earth-analog HZ that could harbor liquid water.}\label{fig:HZ2}
\end{figure}

In this section, the occurrence rate of liquid water bearing planets is estimated using on the probabilistic HZ description. The PDFs of incoming stellar flux and planet radius are taken from \cite{Dressing2013} as discussed in Sect. \ref{sec:PDFs}. That is, we focus on the occurrence rate of HZ planets around M dwarfs here. We do not show the average HZ probability because it is very similar to Fig. \ref{fig:HZ} left. Small differences are caused because the incoming stellar flux and planet radius phase space is non-uniformly covered in log space (see Fig. \ref{fig:KDE}). As a result, some areas of $p_{\mathrm{HZ}}$ are not sampled. The marginalized distributions of $p_{\mathrm{HZ}}$ with respect to the incoming stellar flux and planet radius is shown in Fig. \ref{fig:HZ2}. The marginalized distributions allow us to visualize the uncertainty of $p_{\mathrm{HZ}}$. If we want to convert $p_{\mathrm{HZ}}$ to occurrence rates, we need to multiply the cumulative $p_{\mathrm{HZ}}$ with the total planet occurrence rate of the dataset used. The total planet occurrence rate in the dataset of \cite{Dressing2013} is 1.14 planets/star.

The most optimistic estimate on the occurrence rate is 0.3 habitable planets per star. There are more than 4000 PDF combinations, and thus 4000 estimates on the HZ occurrence rate. The occurrence rate is maximized if water worlds are frequent (optimistic planet mass PDF), rocky planet atmospheres tend to be H$_2$-dominated, and the surface pressure tends to be large, $\sim$10 bars. Under such conditions, a large fraction of small planets (between 0.5 and 2 $R_\oplus$) that receive less than 0.6 $F_\oplus$ could harbor liquid water on the surface. If the incoming stellar flux is larger than 0.6 $F_\oplus$, the fraction of liquid water bearing planets gradually declines to zero at around 2 $F_\oplus$. 

The lower limit on the HZ occurrence rate is zero because we do not know how liquid water is partitioned between the atmosphere, surface, and interior on exoplanets. Atmospheric and geophysical processes could potentially remove all water from the surface. Such processes are e.g., photodissociation of water molecules and the loss of hydrogen to space, the mantle could take up liquid water, or the liquid water can be incorporated into aqueous minerals (e.g., clay). The most pessimistic scenario in the model predicts an occurrence rate of $10^{-5}$ habitable planets per star. In that case, the planet mass follows the pessimistic PDF (ocean planets are rare), the atmosphere is N$_2$-dominated, and the surface pressure is high, $\sim$100 bars.

It is up to future observations to constrain the true distributions of planet and atmosphere compositions, and surface pressure. Until then, the occurrence rate will remain weakly constrained and only optimistic upper limits can be derived from the currently available observational data.

\subsection{The surface pressure and atmospheric composition has the strongest impact on the surface climate}
\begin{figure}
\centering
\includegraphics[width=0.49\textwidth]{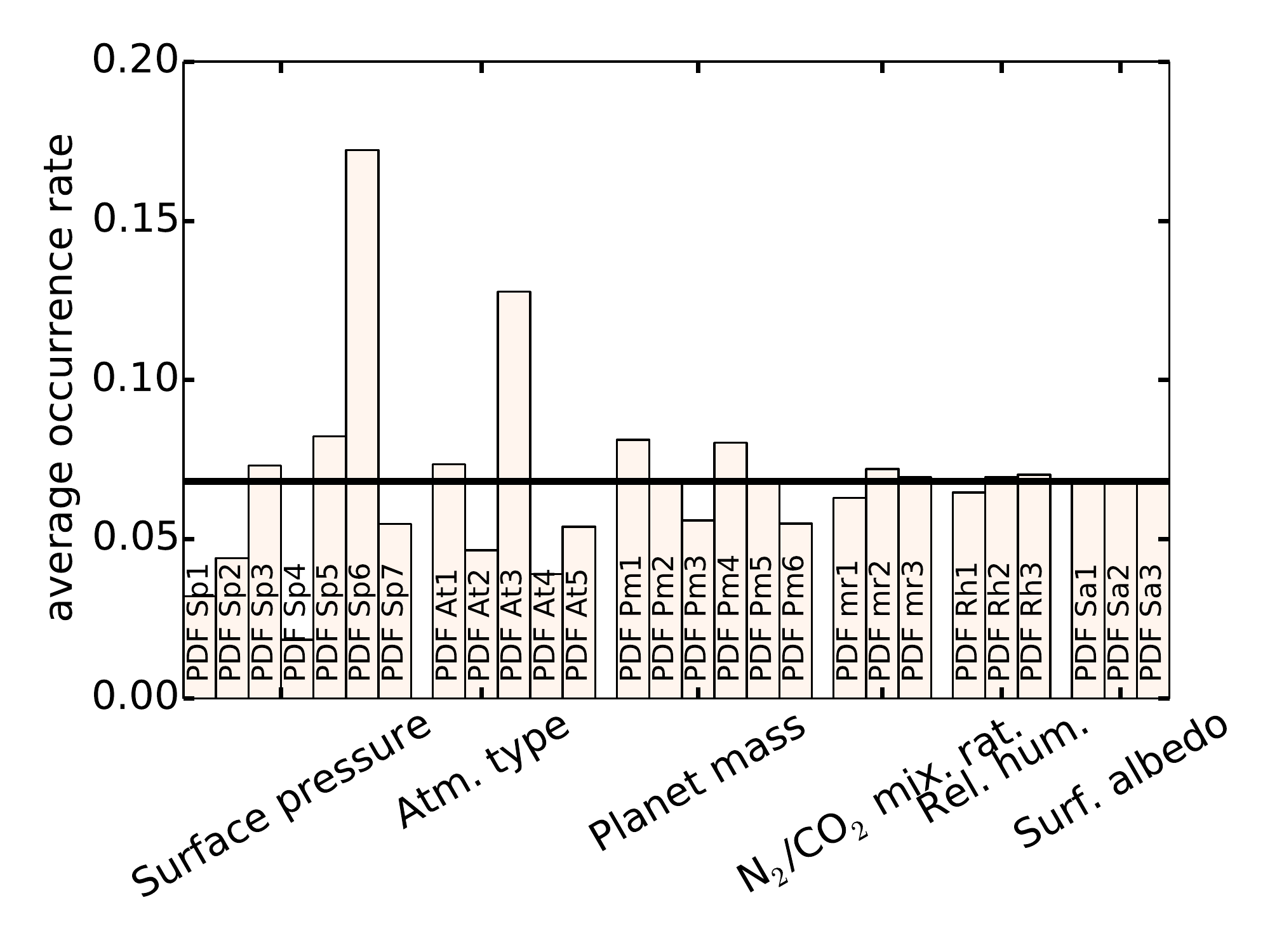}
\caption{The relative importance of unconstrained variables on the HZ occurrence rate estimate. The surface pressure and the atmospheric type have the strongest impact on the HZ occurrence rate. Thus, the HZ occurrence rate would be better constrained if the PDFs of these two variables were observationally known. The rank is established in the following way: There are over 4000 planet populations and thus over 4000 distinct estimates on the HZ occurrence rate. The total average occurrence rate is 0.07 (black solid line). The average occurrence rate for each population that sampled from a given PDF is illustrated by the vertical bars. For example, the average occurrence rate is 0.08 of all populations that sample from the planet mass PDF named `Pm1' (the meanings of the PDFs are described in Tbl. \ref{tbl:PDFs}). If the average occurrence rates of a random variable strongly scatter around black line, the effect of the random variable on the surface climate is strong (e.g., surface pressure and atmosphere type).}\label{fig:impact}
\end{figure}

The observationally unconstrained random variables are ranked based on their impact on the surface climate and the HZ occurrence rate (see Fig. \ref{fig:impact}). Planetary properties that are ranked high on the list have the strongest impact on the surface climate and habitability, and atmospheric characterization efforts should be dedicated to measure these properties to best constrain the habitability of individual planets or planet populations. 

It is possible to rank the random variables because a large number of synthetic populations are available. The first step in ranking is to determine the total average HZ occurrence rate, which is 0.07 planets/star (solid black line in Fig. \ref{fig:impact}). The next step is to determine the average HZ occurrence rate based on those populations that use a given PDF. As all random variables are represented with several PDFs, there are as many HZ occurrence rate estimates as PDFs. If the occurrence rates of one random variable strongly scatter around the total occurrence rate, that random variable has a significant impact on the occurrence rate. 

The most important planet property that has a strong impact on the surface climate is the surface pressure followed by the atmosphere type. The surface pressure is important because pressure broadening of absorption lines can greatly increase the green house effect. The surface pressure of rocky planets and moons in the Solar System vary greatly: 70 bars on Venus, and 0.006 bars on Mars. It is reasonable to assume that the surface pressure on rocky exoplanets is diverse as well, which will have a strong influence on their surface climate. The atmospheric type also has a strong impact on the surface climate. The atmosphere type in the context of the model describes the most abundant component of the atmosphere (i.e., it can be N$_2$, CO$_2$ or H$_2$). Whether the most abundant component is a greenhouse gas or not, has strong implications on the surface climate. 

\section{Discussion}
\label{sec:discussion}

\subsection{Prospects of JWST and the adaptive framework of HUNTER}
We perform a thought experiment in this section to illustrate how additional observational data will help to better constrain the probability that an exoplanet harbors liquid water, and to highlight what will be the greatest challenge in remotely identifying habitable worlds.

Let us assume that we find a transiting planet with a radius of 1 $R_\oplus$ that receives 1 $F_\oplus$ insolation orbiting a nearby M dwarf and the James Webb Space Telescope (JWST) observes its transmission spectrum. Then, we can retrieve the atmospheric composition and the planet mass from the transmission spectrum \citep{deWit2013}. We find that the exoplanet has a mass of 1 $M_\oplus$, a N$_2$-O$_2$-dominated atmosphere with 400 ppm CO$_2$, and some water vapor is also present. That is, the exoplanet appears Earth-like in all aspects. For simplicity, we make the most optimistic assumption that the planet mass and the atmospheric composition are retrieved exactly, without error.

There are three planet properties that could still remain observationally unconstrained: the surface pressure, the relative humidity, and the surface albedo. The surface pressure is unconstrained because we might not be able to distinguish the rocky surface from a cloud deck. Or the surface pressure is larger than a few bars, thus the clear atmosphere is optically thick at all wavelengths and we cannot be sure how deep the atmosphere extends \citep[see Fig. 7 of][]{Benneke2012}. The relative humidity could remain observationally unconstrained for similar reasons. The surface albedo cannot be measured in transmission, only in reflection. And even then, if the surface pressure is too large, light reflected back from the surface might not reach our telescopes.

Under these assumptions, the value of $p_{\mathrm{HZ}}$ can be between 10 - 80\% . As a reminder, we showed in Sect. \ref{subsec:HZ} that $p_{\mathrm{HZ}}$ can be between 0 - 60\% for a planet with 1 $R_\oplus$ that receives 1 $F_\oplus$ around an M dwarf with an otherwise unknown atmospheric composition and planet mass. In other words, $p_{\mathrm{HZ}}$ goes up as expected for an Earth-like atmospheric composition and Earth-like planet mass. However, the uncertainty of $p_{\mathrm{HZ}}$ is still large.

The main reason for the large uncertainty is that the surface pressure is observationally unconstrained. To illustrate this, we fix the surface pressure to be 1 bar but leave the relative humidity and the surface albedo unconstrained. Then, the range of $p_{\mathrm{HZ}}$ becomes 60 - 70\%. The unconstrained surface albedo and relative humidity still introduces some uncertainty, but the range of $p_{\mathrm{HZ}}$ went down significantly.

The thought experiment presented here illustrates that observational and retrieval efforts are paramount to constrain the surface pressure, if we want to identify habitable exoplanets \citep{Misra2014}.

\subsection{1D vs. 3D climate modeling}
\label{sec:1Dvs3D}
A 1D climate model is used in our description, which is a crude approximation of real atmospheres, but it is computationally efficient and allows us to estimate the surface climate for a broad range of planetary and atmospheric parameters. Recently it has been shown that 1D models overestimate the global average surface temperature compared to 3D Global Circulation Models (GCMs) on synchronously rotating planets with 1 bar surface pressure  \cite{Leconte2013,Yang2013}. The discrepancy is significant: while a 1D model predicts 340 K surface temperature and runaway greenhouse conditions, GCMs calculate a global average surface temperature around and below the freezing point of water. The discrepancy exists because 1D models cannot capture the efficient night side cooling.

Even though the shortcomings of 1D models are known, it is computationally unfeasible to generate the look-up table with GCMs. As described in the previous section, the table has seven dimensions for each stellar type considered. If we simulate only three points along each dimension, we need to run 2187 GCM simulations, which is unprecedented to the best of the author's knowledge. Furthermore, the rotation period (or the order of the spin-orbit resonance for tidally locked planets) and the obliquity would be two additional observationally unconstrained variables for the GCM runs. Thus, almost 20000 GCM simulations should be performed if only three points along each axes are considered.

We can estimate under what conditions do we expect the discrepancy between 1D and 3D models to decrease. The key is to consider when heat redistribution is efficient, thus the day-night temperature contrast is small. It has been shown that the day-night temperature difference on synchronously rotating planets (trapped in a 1:1 spin-orbit resonance) becomes smaller if the surface pressure is larger \citep{Joshi1997}. The reason is that the atmospheric relaxation time scale linearly increases with surface pressure \citep{Goody1989}. If the relaxation time scale is large, the atmospheric parcel is transported back to the day side by circulation before it can cool down. The discrepancy is also expected to decrease, if planets are trapped in a spin-orbit resonance that is higher than 1:1. A planet with an initial prograde rotation tends to end up in a high order resonance\footnote{If the resonance is higher order, the rotation and orbital periods are small integer fractions.}, and an initially retrograde planet can reach synchronous rotation \citep{Makarov2012}. A planet trapped in a high order resonance does not have a permanent night side. Thus the day-night temperature difference is expected to be smaller than on synchronously rotating planets. 

As for the impact of 1D modeling on the results, it is expected that the inner edge would be closer to the host star if a GCM or a modified 1D model were used that can treat night side cooling. Atmospheres with $\sim$1 bar surface pressures could still be habitable at incoming flux levels where a 1D model predicts too high surface temperatures. 

\subsection{Correlation between variables}
One caveat of the approach presented here is that I assume the random variables are independent. That is, the value of a random variable does not affect the PDF of other variables. The assumption is justified until our observational knowledge suggests otherwise. It is possible (and straightforward) to include correlations between random variables, but dependencies remain questionable until empirically proven. I highlight two examples here where dependencies might play a role.

It is intuitive to think that the surface pressure should scale with e.g., the planet radius \citep{Kopparapu2014} or potentially with the planet's surface area or mass. Solar System objects with a significant atmosphere do not show such scaling, thus so far we have no evidence to suggest the existence of a trend. If it does exist, it is potentially a weak one with a significant noise. In the future, it might be difficult to find the trend based on observational data. The reason is that the surface pressure of exoplanets can be difficult \citep[but not impossible][]{Misra2014} to constrain because the atmosphere might be optically thick at all wavelengths at high surface pressures, or the atmosphere is cloudy and we observe a cloud deck instead of the surface.

The surface albedo is affected by the surface temperature via the ice-albedo feedback. If the surface temperature approaches the freezing point of water, the surface albedo tends to increase because ice and snow are more reflective. The dependency is expected to have a small influence on the results for two reasons: the ice-albedo feedback is weakened around brown dwarfs \citep{Joshi2012, Shields2013}, and the results suggest that the surface albedo PDFs have a weak influence on the HZ occurrence rate (see Fig. \ref{fig:impact}).

\subsection{Clouds}
The simplest cloud representation was chosen to minimize the number of observationally unconstrained parameters. The radiative effects of clouds are treated as a surface albedo effect. The surface albedo varies between 0 and 0.5 on planets around M dwarfs in agreement with the ice albedo effect \citep{Joshi2012, Shields2013}, and it is between 0 and 1 on planets around solar-like stars, as the surface albedo of a snowball Earth is high around a solar-like star. Clouds are such a complex phenomena, it is likely that a surface albedo approximation remains our best option if exoplanet populations are modeled.

We emphasize that the python implementation of the algorithm (HUNTER) is publicly available. Therefore it can be modified for various cloud treatments in the future\footnote{https://github.com/andraszsom/HUNTER}.

\section{Summary and Conclusions}
\label{sec:sum}
The main challenge in assessing what fraction of exoplanets can potentially harbor liquid water on their surfaces, a crucial ingredient for life as we know it, is that the planet formation history, atmospheric and geophysical properties of exoplanets are currently unknown. Habitability and the occurrence rate of habitable planets can only be weakly constrained without knowledge of the exoplanetary properties because the surface climate is strongly influenced by the atmosphere, geophysics, and the planet formation history. 

The population-based HZ model treats the stellar and planetary properties as random variables, generates a large number of synthetic planet populations, and studies the subpopulation that can harbor liquid water in their surfaces. The stellar and planet properties considered in the model uniquely determine the surface climate of an exoplanet within the limits of a 1D climate model. Current observational data gives information about the stellar properties, the incoming stellar flux that reaches the planet, and the planet radius if we consider transiting exoplanets. The population-based HZ is described as a probability function with respect to these three parameters. That is, the model estimates the probability that an exoplanet harbors liquid water given the host star's spectral type, the incoming stellar flux, and the planet radius.

The main findings are summarized here:
\begin{itemize}
\item Planets that receive more than $\sim$2-3 $F_\oplus$ incoming stellar flux and/or are larger than $\sim$2-3 $R_\oplus$ are unlikely to be habitable. Close in planets receive too much stellar radiation and their surface temperatures are above the boiling point of water. If planets are too large, they are mini-Neptunes without a solid surface. The incoming stellar flux condition means that the HZ has an inner edge. Planets close to the inner edge are ideal for atmosphere characterization with transmission spectroscopy because their transit probability and transit frequency are highest amongst the potentially habitable exoplanets.

\item If the greenhouse effect of the exoplanet's atmosphere is strong, distant planets can maintain habitable conditions on their surfaces. In other words, the outer edge of the HZ could be open. These distant habitable exoplanets can be best characterized with direct imaging in emission in the future.

\item The occurrence rate of HZ planets can only be weakly constrained given our current observational knowledge of exoplanets. Optimistic upper limits can be placed at best because the surface climate of exoplanets is weakly constrained if only the stellar properties, the incoming stellar flux, and the planet radius are known.

\item Out of all observationally unconstrained random variables considered in the model, the surface pressure and the atmospheric composition have the strongest impact on the surface climate of exoplanets. If we want to better constrain the habitability of exoplanets, it is important to know these atmosphere properties.

\end{itemize}

The model developed in this paper focuses on the atmospheric properties of exoplanets and as such, it is concerned with the present day surface climate of the atmosphere, not how it evolved and how long the planet maintains habitable conditions. The argument is that exoplanet characterization with the next generation of telescopes (e.g., JWST, and EELT) will provide constraints on the atmospheric properties of exoplanets. It remains to be seen how well one can infer the geophysical properties and the planet formation history of exoplanets based on their present day atmospheres.

The last two decades of exoplanet sciences were dedicated to the second term in Drake's equation: to observationally constrain the exoplanet population in the Milky Way. The next two decades will be about the third term: what fraction of planets can potentially support life? Liquid water is essential for life as we know it, thus it is important to know what fraction of planets have liquid water on their surfaces. The HZ model outlined in this article will be useful in this endeavor because it captures the expected diversity of exoplanets and it builds on our empirical knowledge. Future observational evidence will hopefully be incorporated into this flexible model to improve the prediction both of the distribution of habitable planets and of the likelihood that a specific planet is habitable, and so improve our understanding of the habitability of the Universe and our chances of finding life there.

\acknowledgements

A. Zsom is grateful for discussions with Sara Seager, Julien de Wit, Vlada Stamenkovic, Stephen Messenger, William Bains, Sascha Quanz, Brice Demory, Felipe Gerhard and the anonymous referee who helped to significantly improve the manuscript. This publication was made possible through the support of a grant from the John Templeton Foundation. The opinions expressed in this publication are those of the authors and do not necessarily reflect the views of the John Templeton Foundation.


\end{document}